\documentstyle[12pt,epsf]{article}

\def\mypagenumber{1}
\def\mydate{October 23, 1995}
\def\myend{\end{document}}

\font\tenrm=cmr10

\normalsize

\newcounter{sxn}

\newcounter{axn}

\date{}

\newdimen\mybaselineskip
\newcommand{\beeq}{\begin{equation}}
\newcommand{\eneq}{\end{equation}}
\newcommand{\beqn}{\begin{eqnarray}}
\newcommand{\eeqn}{\end{eqnarray}}

\def\mybig{\displaystyle \strut }

\def\dd{\partial}
\def\la{\raise.16ex\hbox{$\langle$} \, }
\def\ra{\, \raise.16ex\hbox{$\rangle$} }
\def\go{\rightarrow}

\def\onehalf{ \hbox{${1\over 2}$} }
\def\onethird{ \hbox{${1\over 3}$} }
\def\psibar{ \psi \kern-.65em\raise.6em\hbox{$-$} }
\def\mbar{ m \kern-.78em\raise.4em\hbox{$-$}\lower.4em\hbox{} }

\def\Bbar{ B \kern-.73em\raise.6em\hbox{$-$}\hbox{} }

\def\ep{\epsilon}

\def\vphi{ {\varphi} }

\def\tot{{\rm tot}}

\def\mass{{\rm mass}}

\def\vac{{\rm vac}}

\def\LapN{{\triangle_N^\vphi}}
\def\potN{{V_N(\vphi)}}

\def\myref#1{$^{#1}$}

\def\boxit#1{$\vcenter{\hrule\hbox{\vrule\kern3pt
    \vbox{\kern3pt\hbox{#1}\kern3pt}\kern3pt\vrule}\hrule}$}
\def\bigbox#1{$\vcenter{\hrule\hbox{\vrule\kern5pt
     \vbox{\kern5pt\hbox{#1}\kern5pt}\kern5pt\vrule}\hrule}$}

\begin{document}

\footskip 1.0cm
\thispagestyle{empty}
\setcounter{page}{\mypagenumber}

{\baselineskip=10pt \parindent=0pt \small
\mydate
\hfill \hbox{\vtop{\hsize=3.2cm   UMN-TH-1412/95\\}}
\vspace{6mm}
}

%\vspace*{3mm}

\centerline {\large\bf  Massive multi-flavor Schwinger model}
\vspace*{2mm}
\centerline {\large\bf at finite temperature and on compact
  space\footnote{To appear in the Proceedings of {\it the 4th
International Workshop on Thermal Field Theory and their Application},
Dalian P.R. China, August 5 - 10, 1995.}}
\vspace*{5mm}
\centerline {\tenrm  YUTAKA HOSOTANI}

\vspace*{3mm}

\baselineskip=13pt
\centerline {\small\it School of Physics and Astronomy, University
       of  Minnesota}
\centerline {\small\it Minneapolis, Minnesota 55455, U.S.A.}
\centerline {\small yutaka@mnhep.hep.umn.edu}

\vspace*{3mm}
%\baselinestretch{2.0}
\parindent=12mm
 \small
%\begin{abstract}
\begin{minipage}{12.8cm}
The multi-flavor Schwinger model on $R^1$ at finite temperature $T$
is mathematically equivalent to the model on $S^1$ at $T=0$.
The latter is reduced to a quantum mechanical system of $N-1$ degrees of
freedom.   Physics sensitively depends on the parameter $m/T$.  Finite
temperature  behavior of the massive Schwinger model is quite different
from that of  the massless Schwinger model.
\end{minipage}
%\end{abstract}

\vspace*{3mm}

%\end{titlepage}

%\setcounter{page}{1}

%\textheight=20cm
%\headsep=0.75cm
%\vsize=20cm

%%%%%%%%%%%%%%%%%%%%%%%%%%%%%%%%%%%%%%%%%%%%%%%%%%%%%%%%%%%%%%%%%
\normalsize
\baselineskip=15pt
\parindent=15pt
\vspace*{5mm}

\leftline{\bf 1. Introduction}

\bigskip

The Schwinger model is QED in two dimensions, described by
\beqn
&&{\cal L} = - \hbox{$1\over 4$} \, F_{\mu\nu} F^{\mu\nu} +
\sum_{a=1}^N \psibar_a \Big\{ \gamma^\mu (i \dd_\mu - e A_\mu)
  - m_a \Big\} \psi_a ~~.
  \label{Lagrangian}
\eeqn
Massless theory ($m_a=0$) is exactly solvable.   In the $N=1$ (one flavor)
model the
gauge boson acquires a mass without breaking the gauge
invariance.\myref{1}
It has the $\theta$
vacuum and a non-vanishing chiral condensate $\la \psibar\psi \ra \not=
0$.\myref{2}

The $N\ge 2$  model is distinctively  different from the $N=1$
model.  The spectrum contains $N-1$ massless bosons.  The chiral
condensate vanishes, $\la \psibar\psi\ra=0$,\myref{3} as
 in two dimensions continuous symmetry,
$SU(N)$ chiral symmetry in this case, cannot be spontaneously broken.
Nonvanishing $\la \psibar\psi\ra$ in the $N$=1 theory is allowed, since
the $U(1)$ chiral symmetry is broken by an anomaly.

With massive fermions the model is not exactly solvable.
The effect of the fermion mass in the $N=1$ theory is minor for
$m/e \ll 1$, except that it necessiates the $\theta$ vacuum.
For $N \ge 2$ the situation is quite different.
Coleman showed\myref{4} that in the $N=2$ model with $m_a=m$
two resultant bosons acquire masses given by
\beeq
\mu_1\sim {\mybig \sqrt{2}\, e \over\mybig \sqrt{\pi}}  ~~,~~
\mu_2 \propto  m^{2/3} e^{1/3} \Big| \cos {1\over 2}\theta \Big|^{2/3}~~.
\label{MassColeman}
\eneq
Surprising is the fractional power dependence on $m$, $e$
and $|\cos\onehalf\theta|$ of the
second boson mass $\mu_2$, resulting from the  self-consistent
re-alignment of the vacuum against fermion masses.  The
effect of fermion masses is always non-perturbative even if $m/e \ll 1$.

How does $\mu_2$ depend on $m$ in the $N$ flavor model?  How does it
change at finite temperature?  Does the chiral condensate vanish at
suffieciently high temperature?   Can the fermion mass be treated as
a small perturbation at high temperature?

These are  questions addressed in this article.  We present a powerful
method to evaluate various physical quantities at zero and finite
temperature.  We shall recognize the importance of a dimensionless
parameter $m/T$.  The behavior at $m/T \gg 1$ is quite different from that
at $m/T \ll 1$.  Coleman's result (\ref{MassColeman}) corresponds
to $m/T\gg 1$ as $T\go 0$.  At high temperature $m/T \ll 1$, one finds
$\mu_2 \propto m$, a result obtained in peturbation theory.  This work is
based on the result obtained in ref.\ 5.

\bigskip

\leftline{\bf 2.  At finite temperature and on a circle}
\bigskip

In Matsubara's formalism the model at finite temperature $T$ in
equilibrium is equivalent to an Euclidean field theory, or a theory
with an imaginary time $\tau$, satisfying boundary conditions
\beeq
\psi_a(\tau+{1\over T},x) = - \psi_a(\tau,x)  ~~,~~
A_\mu(\tau+{1\over T},x) =  A_\mu(\tau,x) ~~.
\label{temperatureBC}
\eneq
If one, instead, places the model on a circle
of circumference $L$ (at zero temperature) with boundary conditions
\beeq
\psi_a(t,x+L) = - \psi_a(t,x)  ~~,~~
A_\mu(t,x+L) =  A_\mu(t,x) ~~,
\label{circleBC}
\eneq
then one obtains a theory which is, after Wick's rotation, mathematically
equivalent to the  finite temperature field theory defined by
(\ref{temperatureBC}).  Various physical quantities in the Schwinger model
at $T\not= 0$ are related to corresponding ones in the model on
$S^1$ by substitution of $T$ by $L^{-1}$.

Our strategy is to solve the model on a circle $S^1$ with an arbitrary
size $L$.  There is powerful machinery which specifically works on
$S^1$.\myref{5-8}

\bigskip

\leftline{\bf 3.  Reduction to a quantum mechanical system}

\bigskip

Fermion operators on $S^1$ can be expressed in terms of bosonic operators:
Take $\gamma^\mu = (\sigma_1, i\sigma_2)$ and write
$\psi_a^T=(\psi^a_+,\psi^a_-)$.
In the interaction picture defined by free massless fermions
\beqn
&&\psi^a_\pm(t,x) = { 1\over\sqrt{L}}
\, C^a_\pm \, e^{\pm i \{ q^a_\pm + 2\pi p^a_\pm (t \pm x)/L \} }
  :\, e^{\pm i\sqrt{4\pi}\phi^a_\pm (t,x) } \, :
%&&\framebox[10cm][c]{
%$~~\psi^a_\pm(t,x) = {\mybig 1\over\mybig \sqrt{L}}
%\, C^a_\pm \,
% e^{\pm i \{ q^a_\pm + 2\pi p^a_\pm (t \pm x)/L \} }
%  :\, e^{\pm i\sqrt{4\pi}\phi^a_\pm (t,x) } \, :~~$}
\label{bosonize}
 \\
\noalign{\kern 10pt}
&&\hskip 2cm
e^{2\pi i p^a_\pm} ~ | \, {\rm phys} \ra =  ~ |\, {\rm phys} \ra   ~~~.
\label{phys-cond}
\eeqn
The Klein factors are given by
$C^a_+ =e^{ i\pi \sum_{b=1}^{a-1} ( p^b_+ + p^b_-)}$ and
$C^a_- = e^{ i\pi \sum_{b=1}^{a} ( p^b_+ - p^b_-) }$.
Here $[q^a_\pm, p^b_\pm] = i \, \delta^{ab}$ and
$\phi^a_\pm (t,x) = \sum_{n=1}^\infty (4\pi n)^{-1/2} \,
  \big\{ c^a_{\pm,n} \, e^{- 2\pi in(t \pm x)/L} + {\rm h.c.} \big\}$
where
$[c^a_{\pm,n}, c^{b,\dagger}_{\pm,m}] = \delta^{ab} \delta_{nm}$.
The $:~:$   indicates normal ordering with respect to
$(c_n^{},c_n^\dagger)$.   The antiperiodic boundary condition is ensured by
a physical state condition (\ref{phys-cond}). In
physical states $p^a_\pm$ takes  integer eigenvalues.

After substituting the bosonization formula (\ref{bosonize}),
the total Hamiltonian in the Schr\"odinger picture becomes
\beqn
&&H_\tot = H_0 + H_\phi  + H_\mass \cr
\noalign{\kern 8pt}
&&H_0~~ =  -{e^2 L\over 2} {d^2\over d\Theta_W^2}
 + {\pi\over 2L} \sum_{a=1}^N \bigg\{ (p^a_+-p^a_-)^2
+ (p^a_+ + p^a_- + {\Theta_W\over \pi} )^2 \bigg\}  \cr
\noalign{\kern 8pt}
&&H_\phi = \int_0^L dx \, {1\over 2}  \bigg[ ~
  \sum_{a=1}^N \Big\{ \, \Pi_a^2 + (\phi_a')^2 \Big\}
+{e^2\over \pi}  \Big( \sum_{a=1}^N \phi_a  \Big)^2 ~ \bigg]
    \label{Hamiltonian}
\eeqn
Here $p_+^a - p_-^a$ and $p_+^a + p_-^a$ correspond to the charge and
chiral charge operators, respectively.
$\Theta_W$ is the phase of the Wilson line around the circle, the
only physical degreee of freedom of the gauge field on the circle,
$A_1=\Theta_W(t)/eL$.  The coupling between $p^a_\pm$ and $\Theta_W$ is
induced through the chiral anomaly.\myref{7}
$\phi_a=\phi^a_++\phi^a_-$ and $\Pi_a$ is its canonical conjugate.
$H_\mass$ is the fermion mass term.

Notice that (\ref{Hamiltonian}) is an exact operator identity.
In the absence of fermion masses $H_\tot= H_0 + H_\phi$.
The zero modes ($\Theta_W, q^a_\pm$)
decouple from the oscilatory modes $\phi_a$, and the Hamiltonian is
exactly solvable.  The spectrum  contains one massive field
$N^{-1/2}\sum_{a=1}^N \phi_a$ with a mass $\mu=(N/\pi)^{1/2} e$, and
$N-1$ massless fields.

To examine effects of $H_\mass$,  first note
that  $H_\mass$, and therefore $H_\tot$, commutes with
$p^a_+-p^a_-$.  Hence we can restrict ourselves
to states with $(p^a_+-p^a_-) \, | \, {\rm phys} \ra =0$.  With
this restriction a complete set of eigenfunctions and eigenvalues of
$H_0$ is
\beqn
\Phi^{(n_1, \cdots, n_N)}_s
&=& {1\over (2\pi)^N} \,
   u_s\big[\Theta_W+{2\pi\over N}\sum_a n_a \big] \,
 e^{i\sum n_a  (q^a_++q^a_-)} \cr
E^{(n_1, \cdots, n_N)}_s &=&
\mu (s+\onehalf) + {2\pi\over L} \sum_{a=1}^N  n_a^2
 - {2\pi\over NL} \Big\{ \sum_{a=1}^N n_a \Big\}^2  \label{set1}
\eeqn
where a harmonic oscillator wave function
$u_s$ satisfies $\onehalf (-\dd_x^2 + x^2) u_s = (s+\onehalf) u_s$
with $x = (\pi e^2 L^2/ N)^{-1/4} \Theta_W $.   The ground
states of $H_0$ are  infinitely degenerate for $n_1= \cdots=n_N$ due to
 the invariance under a large gauge transformation
$\Theta_W\go\Theta_W+2\pi$ and $\psi_a \go e^{2\pi ix/L} \psi_a$.

$H_\mass$ induces transitions among $\Phi^{(n_1, \cdots, n_N)}_s $'s.
It also gives finite masses to the $N-1$ previously massless fields.
The structure of the vacuum sensitively depends on $H_\mass$.
The effect of $H_\mass$ turns out
quite nonperturbative so long as $m_a\not= 0$.

It is more convenient to work in a coherent state basis given by
\beeq
\Phi_s(\vphi_a;\theta)
= {1\over (2\pi)^{N/2}} \sum_{ \{ n,r_a \} }
e^{in\theta + i\sum_{a=1}^{N-1}r_a\vphi_a }
  ~ \Phi_s^{(n+r_1, \cdots,n+r_{N-1}, n)} \label{coherent}
\eneq
Transitions in the $s$ index may be ignored to a very good approximation.
We seek  the vacuum in the form
\beeq
\Phi_\vac(\theta) = \int_0^{2\pi} d\vphi_1 \cdots d\vphi_{N-1} ~
   f(\vphi_a;\theta) ~ \Phi_0(\vphi_a;\theta) ~.   \label{vacuum}
\eneq

Let $\chi_\alpha = U_{\alpha a} \phi_a$ and  $\mu_\alpha$ be   a mass
eigenstate field and its mass.   Then matrix elements of
$H_\mass$ in the  coherent state basis are
\beqn
&&\la \Phi_0(\vphi'_a;\theta') | H_\mass | \Phi_0(\vphi_a\theta) \ra
= - \delta_{2\pi}(\theta'-\theta)
 \prod_{b=1}^{N-1}\delta_{2\pi}(\vphi'_b -\vphi_b)
\sum_{a=1}^N  A_a \cos \vphi_a  \cr
&&A_a =  2 m_ae^{-\pi/N\mu L} \,
\prod_{\alpha=1}^N B(\mu_\alpha L)^{(U_{\alpha a})^2} ~~,~~
\vphi_N = \theta - \sum_{a=1}^{N-1} \vphi_a~~.
\label{massMatrix}
\eeqn
$B(z)$ is given by
\beqn
B(z) &=&\exp\bigg\{ \sum_{n=1}^\infty\Big({1\over n}
  -{1\over \sqrt{n^2+(z/2\pi)^2}}\Big) \bigg\}\cr
\noalign{\kern 5pt}
&=& {z\over 4\pi} \exp \bigg\{ \gamma + {\pi\over z}
 - 2 \int_1^\infty  {du \over (e^{uz} - 1)\sqrt{u^2-1}}  \bigg\}~.
\label{massOperator}
\eeqn
%Note that $B(z) \sim ze^\gamma/4\pi$ for $z\gg 1$, and $B(0)=1$.
The eigenvalue equation $(H_0+H_\mass) \, \Phi_\vac(\theta) = E\,
\Phi_\vac(\theta) $ becomes
\beqn
&&\Big\{ -\LapN + \potN \Big\} ~ f(\vphi)
= \ep~ f(\vphi) \cr
\noalign{\kern 12pt}
&&\hskip 1cm \LapN =
\sum_{a=1}^{N-1} {\dd^2\over \dd\vphi_a^2}
-{2\over N-1} \sum_{a<b}^{N-1} {\dd^2\over \dd\vphi_a \dd\vphi_b} \cr
\noalign{\kern 5pt}
&&\hskip 1cm \potN = -{NL\over 2(N-1)\pi}
{}~  \sum_{a=1}^N  A_a  \cos \vphi_a     ~~.
 \label{QMeq}
\eeqn
Here $\ep=NEL/2\pi(N-1)$.  Eq.\ (\ref{QMeq}) is nothing but the
Schr\"odinger equation with the kinetic and potential terms given by
$-\LapN$ and $\potN$, respectively.

The potential $\potN$
depends, through $A_\alpha$ defined in (\ref{massMatrix}), on
$\mu_\alpha$ and $U_{\alpha a}$ which are to be self-consistently
determined from the ground state wave function $f(\vphi_a;\theta)$
of the Schr\"odinger equation (\ref{QMeq}).
 $\mu_\alpha$'s and $U_{\alpha a}$'s are determined by
\beqn
&& {\mu^2\over N} \pmatrix{ 1 &\cdots & 1 \cr
                                     \vdots & \ddots & \vdots\cr
                                     1 & \cdots & 1 \cr}
  + \pmatrix{ R_1 &&\cr
              &\ddots&\cr
              &&R_N\cr}
= U^T \, \pmatrix{ \mu_1^2 &&\cr
              &\ddots&\cr
              &&\mu_N^2\cr} \, U    \cr
\noalign{\kern 10pt}
&&R_a = {4\pi\over L} \,  A_a  \la \cos \vphi_a \ra_f
 = - 4\pi m_a \, \la \psibar_a\psi_a \ra_\theta~~.
\label{diagonalizeMass}
\eeqn
We have denoted  $\la g(\vphi)\ra_f=\int [d\vphi]\, g(\vphi) |f(\vphi)|^2 $.
We need to solve (\ref{massMatrix}), (\ref{QMeq}),  and
(\ref{diagonalizeMass}) self-consistently.

We have shown that the
$N$ flavor massive Schwinger model is reduced to  quantum mechanics
of $N-1$ degrees of freedom in which the potential has to be fixed
self-consistently with its ground state wave function.

\begin{figure}[tb]
\epsfxsize= 8.cm    % changed from 10 cm  to 8.5 cm.  9cm is too big.
\epsffile[60 120 410 650]{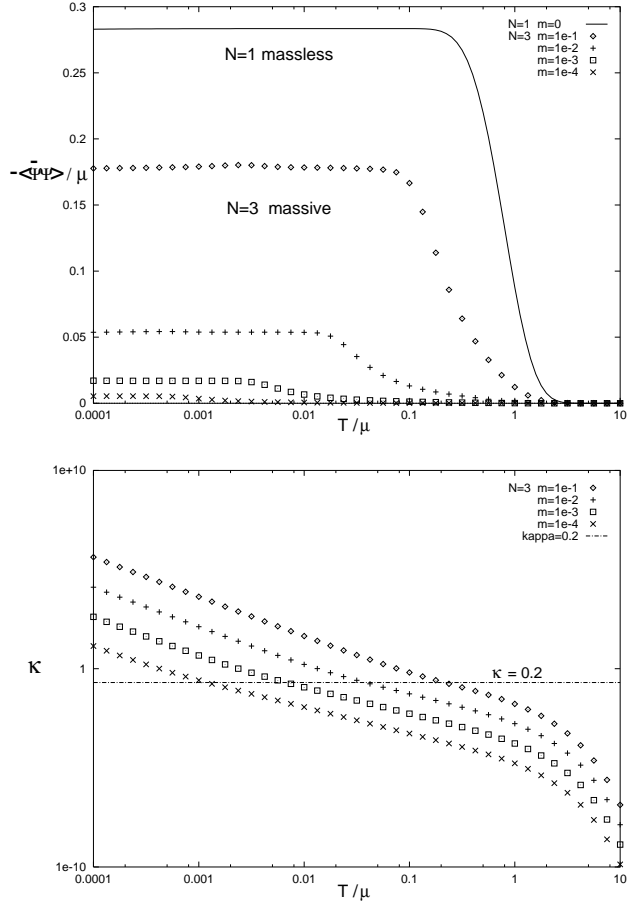}
%\epsffile[120 120 470 650]{T-chi-kappa.ps}   %---10/20
\vskip -0cm
\caption{Chiral condensate $-\la \psibar\psi\ra/\mu$ and $\kappa_0$ as a
function of temperature $T$ at $\theta=0$ in the $N=1$ and $N=3$ models.
In the $N=3$ model $m/\mu = 10^{-1}, 10^{-2}, 10^{-3}$ and $10^{-4}$.
The crossover in $\la \psibar\psi\ra/\mu$ takes place when $\kappa_0 \sim
0.2$.}
\label{fig:1}
\vskip 0.3cm
\end{figure}

%\bigskip
\newpage
\leftline{\bf 4.  N=1 (one flavor)}

\bigskip

One flavor case is special.  $\vphi_1=\theta$ and there is only one
massive boson with a mass $\mu_1$ .  This case was analysed in detail in
refs.\ 7 and 9.

The vacuum on $S^1$ is $\Phi_\vac(\theta) = \Phi_0(\theta)$.
Converting all expressions to finite temperature case, we find the
boson mass and chiral condensate to be
\beqn
&&\mu_1^2 = \mu^2 + 8\pi mT \, \cos\theta\, e^{-\pi T/\mu} \,
   B\Big({\mu_1\over T}\Big) \cr
\noalign{\kern 5pt}
&&\la\psibar\psi\ra_\theta =
-2T \, \cos\theta\, e^{-\pi T/\mu} \,  B\Big({\mu_1\over T}\Big)
\label{condensateN=1}
\eeqn
where $\mu=e/\sqrt{\pi}$.  At $T=0$
\beqn
&&\mu_1 = \sqrt{\mu^2 + m^2e^{2\gamma}\cos^2\theta} +
me^\gamma\cos\theta\cr
\noalign{\kern 5pt}
&&\la\psibar\psi\ra_\theta = - {e^\gamma\over 2\pi} \, \mu_1 \cos\theta~.
\label{condensateN=1-T=0}
\eeqn

Notice that the condensate is nonvanishing even for $m=0$.
As $T$ is raised,  it  shows a crossover around $T=\mu$,
and approaches zero at high temperature.  See fig.\ 1.
The correction due to the fermion mass $m\ll \mu$ is minor.

\bigskip

\leftline{\bf 5. N$\ge$2 (multi-flavors) at low  and high T}

\bigskip

The situation in the multi-flavor case is quite different.  When
$m_a=0$, $\potN=0$ and $f(\vphi)=$const so that $\la \cos\vphi_a\ra_f =0$.
Consequently $\la \psibar_a\psi_a\ra_\theta=0$.

Suppose that fermion masses are degenerate: $m_a=m$.  There results
one heavy boson and $N-1$ light bosons with masses $\mu_1$ and $\mu_2$,
respectively.  The potential in (\ref{QMeq}) becomes
\beqn
&&\potN = ~-  \kappa_0 ~  \sum_{a=1}^N \cos \vphi_{a} \cr
\noalign{\kern 10pt}
&&\kappa_0 = {N\over (N-1)\pi} ~ {m\over T}
    ~B\Big({\mu_1\over T}\Big)^{{1\over N}}
     B\Big({\mu_2\over T}\Big)^{1-{1\over N}} ~ e^{-\pi T/N\mu}
   \label{potential1}
\eeqn
where $\mu_1$ and $\mu_2$ are determined by
\beqn
&&\mu_1^2 = \mu^2 + \mu_2^2\cr
\noalign{\kern 12pt}
&&\mu_2^2 =  {8\pi^2 (N-1)\over N} ~ \kappa_0 T^2\,
    \la \cos \vphi \ra_f  = - 4\pi m \la \psibar_a\psi_a\ra_\theta ~~.
   \label{masses1}
\eeqn
Recognize that two parameters $\kappa_0$ and $\theta$ fix the
potential $\potN$.

If $m\not=0$, $\kappa_0$ becomes very large at low temperature
$T\go 0$.  In this regime the potential term dominates over the
kinetic energy term in the Schr\"odinger equation (\ref{QMeq}).
The wave function $f(\vphi)$ has a sharp peak around the minimum of
the potential.  The minimum is located at $\vphi_a = \bar\theta/N$
where $\bar\theta= \theta - 2\pi[(\theta/2\pi)+\onehalf]$.  As $\theta$
varies from $-\pi$ to $+\pi$, the minimum moves from $\vphi_a=-\pi/N$
to $\vphi_a=+\pi/N$, and jumps back to $\vphi_a=-\pi/N$.

In the $T\go 0$ limit, $ \la \cos \vphi \ra_f  = \cos\bar\theta$ so that
\beeq
{1\over \mu} \, \la \psibar\psi \ra_\theta
= - {1\over 4\pi} \Big( 2e^\gamma \cos {\bar\theta\over N} \Big)^{{2N\over
N+1}}  \, \Big({m\over \mu}\Big)^{{N-1\over N+1}}
   \hskip .5cm {\rm for} ~~ T \ll m^{N\over N+1} \mu^{1\over N+1}~.
\label{condensateN-T=0}
\eneq
Two important observations follow.  First, the dependence of
the condensate   on $m$ is non-analytic.  It has fractional power
dependence.   The effect of fermion masses is nonperturbative in this
limit.  Secondly, as a function of $\theta$, the condensate has a cusp
at $\theta=\pm\pi$, which originates from the discontinuous jump
in the location of the minimum of the potential.

In the opposite limit $\kappa_0 \ll 1$, which includes both $m\go 0$
(with $T>0$ kept fixed) and $T\gg \mu$, the potential $\potN$ can be
treated as a small peturbation in (\ref{QMeq}).  One finds
\beeq
\la \cos \vphi_a \ra_f
=\cases{(1+\cos\theta) \kappa_0  &for $N=2$\cr
        \kappa_0  &for $N\ge 3$.\cr}   \label{smallKappa}
\eneq
There appears no $\theta$-dependence for $N\ge 3$  to this
order.  The condensate for $N\ge 3$ is found to be
\beeq
{1\over \mu}\, \la \psibar\psi \ra_\theta
   = - {2N\over \pi(N-1)} \, { m\over \mu} ~
\cases{
 \bigg( {\mybig \mu e^\gamma \over\mybig 4\pi T} \bigg)^{2/N}
  &for $m^{N\over N+1} \mu^{1\over N+1} \ll T \ll \mu$\cr\cr
  e^{-2\pi T/N\mu} &for $T \gg \mu$\cr}
  \label{condensateN-highT}
\eneq
For $N=2$, the expressions for  $\la \psibar\psi \ra_\theta$   in
(\ref{condensateN-highT}) must be multiplied by a factor
 $2 \cos^2 \onehalf\theta $.  Notice that the condensate is
linearly proportional to $m$ in this regime.

\begin{figure}[bt]
\epsfxsize= 12cm
\epsffile[0 230 500 580]{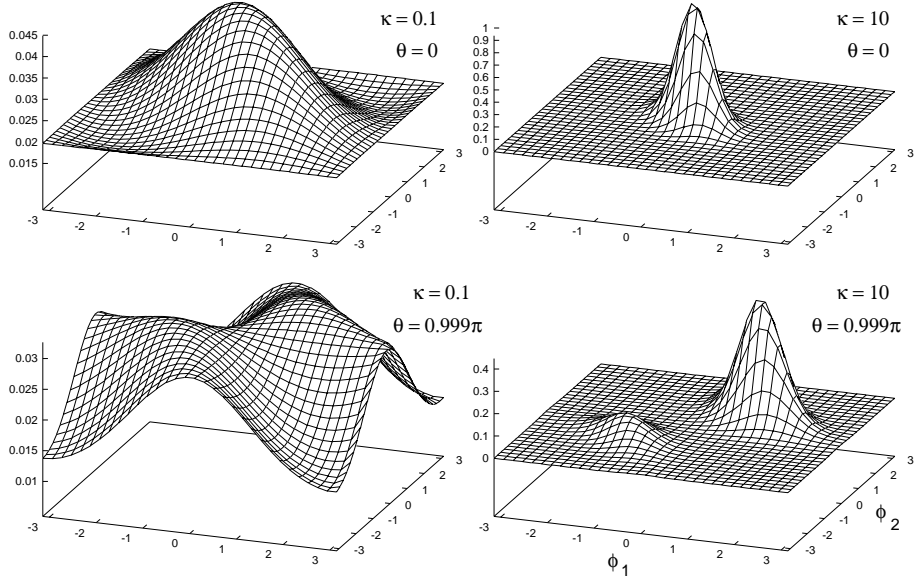}
\vskip 0cm
\caption{\tenrm Wave function $|f(\vphi)|^2$ in the $N=3$ model.}
\label{fig:2}
\vskip 0.2cm
\end{figure}

\bigskip
%\newpage
\leftline{\bf 6. Crossover in $\la \psibar\psi \ra_{\theta,T}$}
\bigskip

For general values of $T$ and $m$, we need to solve the set of equations
numerically.  In the two flavor case the Schr\"odinger equation is
\beeq
\Bigg\{ - {\dd^2\over \dd\vphi_1^2} - 2 \kappa_0 \cos\onehalf\theta
 \cos(\vphi_1-\onehalf\theta) \Bigg\} \, f(\vphi_1)
 = \ep \, f(\vphi_1) ~.
 \label{N=2eq}
\eneq
This is the equation for a quantum pendulum.  The strength of the
potential is $2\kappa_0\cos\onehalf\theta$, which changes
the sign at  $\theta=\pm\pi$.   In particular, the potential vanishes
at $\theta=\pm\pi$ and $f(\vphi_1)=$constant. Consequently the chiral
condensate vanishes at $\theta$=$\pm\pi$.  This is a special feature of $N=2$.

For $N=3$ the equation is
\beeq
\Bigg\{ -{\dd^2\over \dd\vphi_1^2}
        -{\dd^2\over \dd\vphi_2^2}
        +{\dd^2\over \dd\vphi_1 \dd\vphi_2}
- \kappa_0 \Big[\cos\vphi_1 + \cos\vphi_2
  + \cos(\vphi_1+\vphi_2-\theta) \Big]   \Bigg\}  ~ f(\vphi)
   = \ep~ f(\vphi) ~~.
 \label{N=3eq}
\eneq
The potential term never vanishes unless $\kappa_0=0$, or equivalently
$m=0$ or $T\go\infty$.  The equation can be solved numerically
for an arbitrary $\kappa_0$.  The ground state wave function
$|f(\vphi)|^2$ has been displayed for various $\kappa_0$ and $\theta$ in
fig.\ 2.

With given values of $m$ and $T$
the chiral condensate is determined by solving (\ref{potential1}),
(\ref{masses1}), and (\ref{N=3eq}) simultaneously.  We developed an
iteration procedure which yields a consistent set of values of
$m/\mu$, $T/\mu$, $\mu_a/\mu$, and $\kappa_0$.  It takes less than ten
iterations even for moderate values of $\kappa_0\sim 1$ to achieve
four digits accuracy.

In the top figure of fig.\ 1 we have displayed the condensate
$\la\psibar\psi\ra_{\theta=0}/\mu$ as a function of $T/\mu$ and $m/\mu$ in the
$N$=3 case too.
With a given $m/\mu$ the condensate is almost constant at low $T/\mu$, and
sharply drops to a small value around $T_*$.  This crossover takes place
when $\kappa_0(m,T_*) \sim 0.2$ for a wide range of the value of $m/\mu$.  In
the bottom figure we have shown a plot for $\kappa_0$ in the same region of the
$m$-$T$ space.

The asymptotic formulas (\ref{condensateN-T=0})
and (\ref{condensateN-highT}) are quite accurate for $\kappa_0 >1$
and $\kappa_0 <0.1$.  The important parameter is $\kappa_0$.

\begin{figure}[tb]
\epsfxsize= 10cm
\epsffile[50 190 470 500]{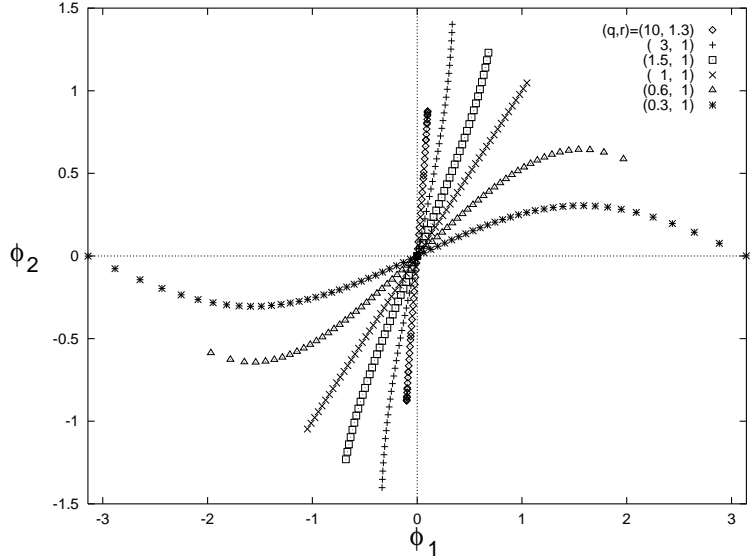}
\vskip 0cm
{\small
\caption{The location of the minimum of the potential in the $N=3$
model is displayed with various values of $(q,r)$ in (23).
For $(q,r)$=(10, 1.3) the minimum at $\theta$=$\pm\pi$ is located at
the origin.}
}
\label{fig:3}
\vskip 0.3cm
\end{figure}

\bigskip
\bigskip
\leftline{\bf 7. The singularity at $\theta=\pm\pi$ and fermion masses}
\bigskip

As shown in (\ref{condensateN-T=0}), the chiral condensate at $T=0$
shows a cusp singularity in its $\theta$ dependence when  fermion
masses are degenerate.  When fermion masses are not degenerate, the
potential $\potN$ in (\ref{QMeq}) is deformed from the symmetric one
in (\ref{potential1}).  Accordingly the location of the minimum
is shifted.  The cusp singularity appears when the location of the
minimum of the potential makes a discontinuous jump.

With given $\{ m_a \}$, $A_a$ in $\potN$ is determined
self-consistently.  We have determined the location of the minimum of
the potential in the three flavor case.  The potential is proportional
to
\beeq
F(\vphi_1,\vphi_2;\theta) =
-q\cos\vphi_1 - r \cos\vphi_2 - \cos(\theta-\vphi_1-\vphi_2)
\label{potential2}
\eneq
where $q=A_1/A_3$ and $r=A_2/A_3$.

In the symmetric case $q$=$r$=1
the location of the minimum moves from $(-\onethird\pi, -\onethird\pi)$
to  $(+\onethird\pi, +\onethird\pi)$ as $\theta$ varies from $-\pi$ to
$+\pi$, and makes a jump.  At $\theta$=0 the minimum is located at the
origin for arbitrary $(q,r)$.  We have plotted trajectories of the
location of the minimum for several typical values of $(q,r)$ in
fig.\ 3.

So long as asymmetry is small, there is a discontinuous jump at
$\theta=\pm \pi$.  However, a sufficiently large asymmetry restores
continuity.  For instance, with $(q<0.5, r=1)$ the minimum at
$\theta=\pm\pi$ is located at $(\vphi_1,\vphi_2)=(\pm\pi,0)$.  The
trajectory makes a closed loop on the $\vphi_1$-$\vphi_2$ torus.
For $(q\gg 1,r=1)$, the minimum at $\theta=\pi$ is around
$(0,\onehalf\pi)$ so that the discontinuity remains.  However,
if one adds a small aysmmetry in ``$r$'', the minimum is pushed back
to the origin.  For instance, for $(q,r)=(10,1.3)$,  the minimum starts
to turn back at $\theta \sim 0.8\pi $ and reaches the origin at $\theta=\pi$.
The implication to QCD physics is profound.  As $m_s \gg m_d > m_u$,
we are facing at a case $q\gg r > 1$ in which there is no
singularity in $\theta$ any more.\myref{10}

We conclude that a sufficiently large asymmetry in the fermion masses
removes the cusp singularity at $\theta=\pm\pi$ in
$\la \psibar\psi \ra$ at $T=0$.

\vskip .5cm
\leftline{\small\bf Acknowledgments}

{\small
This work was supported in part by by the U.S.\ Department of Energy
under contracts   DE-AC02-83ER-40105.   The author  would like to thank the
Aspen Center for Physics  for its hospitality where a part of the
work was carried out.
}

\vskip .5cm

\def\ap {{\it Ann.\ Phys.\ (N.Y.)} }
\def\cmp {{\it Comm.\ Math.\ Phys.} }
\def\ijmpA {{\it Int.\ J.\ Mod.\ Phys.} {\bf A}}
\def\ijmpB {{\it Int.\ J.\ Mod.\ Phys.} {\bf B}}
\def\jmp {{\it  J.\ Math.\ Phys.} }
\def\mplA {{\it Mod.\ Phys.\ Lett.} {\bf A}}
\def\mplB {{\it Mod.\ Phys.\ Lett.} {\bf B}}
\def\plB {{\it Phys.\ Lett.} {\bf B}}
\def\plA {{\it Phys.\ Lett.} {\bf A}}
\def\nc {{\it Nuovo Cimento} }
\def\npB {{\it Nucl.\ Phys.} {\bf B}}
\def\pr {{\it Phys.\ Rev.}}
\def\prl {{\it Phys.\ Rev.\ Lett.}}
\def\prB {{\it Phys.\ Rev.} {\bf B}}
\def\prD {{\it Phys.\ Rev.} {\bf D}}
\def\prp {{\it Phys.\ Report} }
\def\ptp {{\it Prog.\ Theoret.\ Phys.} }
\def\rmp {{\it Rev.\ Mod.\ Phys.} }
\def\hep {{\tt hep-th/}}

\bigskip
%\newpage
\parindent=0pt
\parskip=0pt
{\bf  References}
\begin{enumerate}
\small
\baselineskip=13pt
\parsep=0pt
\listparindent=0pt
\parskip=0pt
\item
 J. Schwinger, \pr {\bf 125} (1962) 397;  {\bf 128}  (1962) 2425;
\item
J.H. Lowenstein and J.A. Swieca, \ap {\bf 68}  (1971) 172;
 A. Casher, J. Kogut and L. Susskind, \prD {\bf 10} (1974) 732;
S. Coleman, R. Jackiw, and L. Susskind,  \ap {\bf 93} (1975) 267.
\item
M.B. Halpern, \prD {\bf 13}  (1976) 337;
I. Affleck, \npB {\bf 265} [FS15] (1986) 448.
\item S. Coleman,  \ap {\bf 101}  (1976) 239.
\item J.E. Hetrick, Y. Hosotani, and S. Iso, \plB {\bf 350} (1995)  92;
{\it `The interplay between mass, volume, $\theta$, and
$\la\psibar\psi\ra$ in $N$-flavor QED$_2$'}, {\tt hep-th/9510090}.
\item  N. Manton, \ap {\bf 159} (1985)  220;
\item J.E. Hetrick and Y. Hosotani, \prD {\bf 38} (1988) 2621.
\item
 R. Link, \prD {\bf 42}  (1990) 2103;
 S. Iso and H. Murayama, \ptp {\bf 84}  (1990)142.
\item
 I. Sachs and A. Wipf, {\it Helv. Phys. Acta.} {\bf 65} (1992) 652.
\item Creutz,  {\tt hep-th/9505112}.

\end{enumerate}

%%%%%%%%%%%%%%%%%%%

\myend